# 9.5 Hypotheses on the Informational Structure of Life which are Multi-scale, Human-readable, Provocative, Coherent, Plausible, Falsifiable, and Actionable

William Softky
P.O. Box 872
Menlo Park, CA 94025
December 28, 2017

*In memory of my brother and collaborator Ed Softky, November 7 1963 – October 9 2008, who wanted to teach Buddhist Principles through software concepts.*

**Abstract**

This document has an unorthodox structure to accomplish an unorthodox goal: presenting the scaffolding for a zero-parameter Unified Theory of Life inside several dozen pages. Ideally, that length is short enough to be understood in a few hours, yet complete and principled enough to unfold into solutions to civilization's most urgent problem: the accelerating de-calibration of fluid human brains by compelling digitized signals. The stakes of understanding humans as informational beings are huge. Without drastic changes, crucial human functionality will vanish in a generation. Fortunately, the cures are cheap and easy if promoted properly. These ten hypotheses re-present in more abstract, encapsulated form a Framework published a few months ago in a reputable peer-reviewed journal, which concluded that human sensory systems must be accorded data of the same quality as the data which already trains algorithmic intelligences. While the lack of experiment-grade detail might make these hypotheses seem "unscientific," in compensation the breadth of this Framework ought to provide the virtues of theory: clarity, simplicity, coherence, and self-evidence. Five of the hypotheses span humanity's current problem-space, and five a possible solution space.  They are:
H1.0 Stably evolving distributions must balance themselves between narrowing and broadening
H2.0 Stabilization and homeostasis are fragile in multiple ways
H3.0 Representing spacetime requires micro-timing and mega-assumptions
H4.0 When in doubt, ping!
H5.0  Mediated communication becomes infected with pinging
H6.0 If scaling and incentives are the problems, then entropy and affection are the solutions
H7.0 "Paleo everything but violence" provides sensorimotor nutrition
H8.0 Humans evolved to resonate ecstatically
H9.0 Symmetric spinal health syndrome
H9.5 Helping the irresistible force beat the immovable object

## [Meta-data about this paper]

This document has an unorthodox structure to accomplish an unorthodox goal: presenting the scaffolding for a zero-parameter Grand Unified Theory of Life inside a couple dozen pages.  To keep this paper short enough, I choose the most condensed scientific communication channel possible: coherent, testable hypotheses stripped of ornament.  While necessary here, such compression renders many typical features of journal articles irrelevant, such as equations, data, references (recent or otherwise), and specific experimental predictions.

Ideally, this paper is short enough to be understood in a few hours, yet complete and principled enough to help solve civilization's most urgent problem: the accelerating de-calibration of fluid human brains by digitized signal environments and over-compressed communication. This year even more than last, mental and political health worldwide suffer from our failure to understand ourselves as lifeforms with informational needs.

The stakes involved in understanding ourselves as informational beings are huge. I have recently become co-author on a conventional, peer-reviewed paper which proposed a Framework for that process. That Framework explains a host of digital dependencies (so-called "addictions" to web-surfing, texting, gaming addiction, or social media) as the nearly inevitable result of feeding continuous nervous systems discrete training data. Those startling conclusions are already influencing public debates. If in fact digital influences are responsible for the unprecedented year-on-year worldwide increase in childhood mental illness and suicide, then the next generation of human beings may be socially incapable of even nurturing their own children. A one-generation catastrophe would make global warming look slow. A crucial component of humanity would be lost, perhaps forever.

Fortunately, the cures are cheap and easy if promoted properly. Informational needs allow informational cures—roughly, continuous vibratory interaction with a variety of organic signal-sources—which as a bonus tend to deliver deeply satisfying forms of pleasure. Once people know what to do, they like doing it.

The Framework underlying the present Hypotheses is the same as the one formally published a few months ago. The purpose of the present paper is to meld the knowledge from two deeply principled forms of science: the traditional physical/thermodynamic understanding of the world ("physics"), and the signal-processing/computational/mathematical/AI/thermodynamic understanding ("information theory").

Toward that goal, many disciplines such as computational physics, cybernetics, neural nets, nonlinear dynamics, complexity theory, and so on have already built on this basic equation:

**Physics + Information = Life**

The present approach differs principally by beginning with geometric principles, such as representation of 4-D spacetime and the narrowing or widening of probability curves. Among many results, this approach concludes that a brain's primary function is *continuous high-precision representation of spacetime*, with correspondingly strong training-data and micro-timing requirements on input and interactions.

A casual reader might conclude this paper's lack of traditional features of journal articles (equations, data, references, experimental predictions) makes these hypotheses "unscientific." Those more familiar with the long-term function of the scientific method will recognize the crucial role of undisputed scientific principles in coordinating general research agendas, and the role of well-crafted specific hypotheses in validating them.

As a matter of principle, the level of specificity necessary to publish in a single discipline could never in fairness be also required when unifying many disciplines at once.  Equation-grade and experiment-grade detail, while appropriate for the vast majority of scientific research, necessarily distract from understanding basic principles, like gravitation and entropy, which operate at all scales at once. If I could find a pre-existing presentation format, I would copy it, but there is no "Journal of Grand Unified Theories"—especially for Theories of Life!—to show how the present theory should be formatted at the desired compression level. So, like the theory itself, the format of this document must be derived from first principles.

No single equation could possibly cover all Life's forms of self-replication, from DNA up to blueprints and legalese. No specific data could support any claim of universality, as could no single reference (even background references; the necessary concepts are already in common circulation).  No recent results save information theory, physics, and geometry are necessary, since such Laws ought to be eternal, and be relevant both to human informational needs and to the informational structure of Life.

And, because this document proposes whole classes of testable experiment, specifics would be superfluous.  In summary: no equations, no data, and no experimental tests in this paper. Just testable hypotheses, and the many questions they pose.

What merits might this paper offer to replace the usual granular detail?  The usual virtues of theory: clarity, simplicity, coherence, and self-evidence. Any unification should offer a single language potentially in common with all applicable disciplines, and should appeal to geometric intuition, as Newton's Laws do. Grand Science must necessarily be scored differently than small science, just as theory is scored differently than experiment. For Science to enjoy its fundamental unity, all those regimes are necessary, so even a poor unification is worth setting up at first as a straw man, to spur stronger works. This paper might be that straw-man.
The introductory section **Warrants** summarizes basic geometrical and physical principles invoked as axioms supporting the hypotheses (often in terms from software and signal-processing, like *bandwidth* and *interface format*, which best serve informational claims).  At the beginning only the concepts are listed, with  full-text explanation in Appendix A.

Of the hypotheses themselves, the first four span the basic functions of life: self-replication, self-regulation (including foraging), representation of spacetime, and self-calibration. The fifth hypothesis describes a virulently catastrophic failure mode of self-calibration which may explain the recent surge in mental illness worldwide. The final four hypotheses address solutions to this catastrophe through specific, optimal structures of human organization, sensorimotor interaction, interpersonal mechanical resonance, and intra-personal (spinal) neuromechanical resonance. The final half-hypothesis is a call to action.  Each is described at four levels of detail. The abstract gives only the hypotheses' brief titles.  The introduction provides those, a single detailed sentence and a page or so of description each, along with some research questions. Appendix B enumerates those same questions further.

Many remarkable possibilities are implied by—but not described in!—these hypotheses. For example, there are likely to exist neuromechanical self-help tricks by which an individual might shed life-long muscular pain and stiffness in weeks, or a disaffected couple might immediately experience resonant neuromechanical ecstasy in hours. The mechanisms are the same as those of Yoga, Pilates, Feldenkrais etc., but now geometrically distilled.

Each hypothesis, if true, provides only a single bedrock of understanding. I hope that the nine together can serve as a continuous path of stepping-stones from incontestable eternal truths about the informational structure of life on the one hand to immediate solutions of real human problems on the other. In other words, that these separate ideas nonetheless implement not just the goal of "Natural Philosophy," but the goal of Philosophy itself, by connecting the scientific truth of "What is?" directly to the ageless human question, "How then shall we live?"

**[Content: Warrants and Hypotheses]**

**Warrants: The Geometric Laws of Nature**

These concepts listed below contain foundational assumptions of these hypotheses. Their explanation and context is in Appendix A.

> entropy & information
> spacetime
> architectural layer
>  no boundary condition
> continuous
>  multi-scale
> symmetries: translation, rotation, dilation
> reference frames: spherical, cylindrical, Cartesian
> Platonic solids
> twist, shear, dilate, break, bend, expand
> software inheritance
> fractures
> seamless, continuous, connected
> continuous interface format
> infinite fine-ness (resolution)
> blurred, uncertain
> discrete, distinct
> combinatorial
> continuous, low-dimensional Laws like temperature and pressure
> compress …into a couple scalar parameters
> Shannon's Laws of Information
> aggregates
> simple, low-entropy descriptions
> principles, evidence
> gathering data, interpreting data, planning action
>  "What is?"
> "How then shall we live?"
> Root Principles
>  energy sources
>  entropy sinks
> non-equilibrium thermodynamics
> active stabilization
> homeostatic
>  (meta)stable self-regulation
> resource allocation strategy (energy, time, location)
> parameter space

blurred, broadly distributed
specialized, narrow, focused
distribution, central value, populations
sharpening forces
entropy-reducing mechanism
self-replication, , selection, amplification, edge-enhancement, specialization
specific physical mechanism
abstract transformation applied to a probability-distribution function
entropic consequences
algorithmic ease
data-constraints of stabilized systems
Occam's Razor
algorithmic training
statistical inference
high-dimensional, complex spaces
Curse of Dimensionality
high-entropy, variable, multi-resolution data
natural, naturalistic inputs
low-entropy "test-patterns"
fractured models
over-fitting

# Hypotheses regarding the informational structure of Life

## H1.0 Stably evolving distributions must balance themselves between narrowing and broadening

**In order for any abstract distribution to persist over time, the forces broadening vs. sharpening its distribution profile must be in balance, so that the passive blurring forces of diffusion, dispersion, and mutation must be actively counterbalanced by sharpening forces of amplification and selection; one particular mechanism could be that growth among the distribution's lower-entropy components raises the overall entropy density, while growth among the higher-entropy components decreases it.**

Entropy density describes the complexity of patterns along the widest-possible range, from near zero entropy (strictly structured, standardized, and regimented, like a crystal) to a maximum, fully atomized and independent. Imagine an ensemble of information-processing actors, creating changing patterns as they act and interact. They could be strands of DNA, phrases of legalese, memes, or re-copied images. They exist in some physical medium, so the ensemble as a whole has an "entropy density" $\rho_H$. Decreasing entropy density means the system is becoming more structured; increasing means it is becoming more diverse. Life patterns fall in between, with the largest and more complex structures (animals and societies) being more crystal-like, and small independent ones being more gas-like. One could plot the population-success of any ensemble along that axis, with the most self-sharpening ensembles toward the left and the most quickly diffusing ones at right. At left is the realm of "sharpening traps," in which over-sharpening creates over-brittle, nearly singular structures.

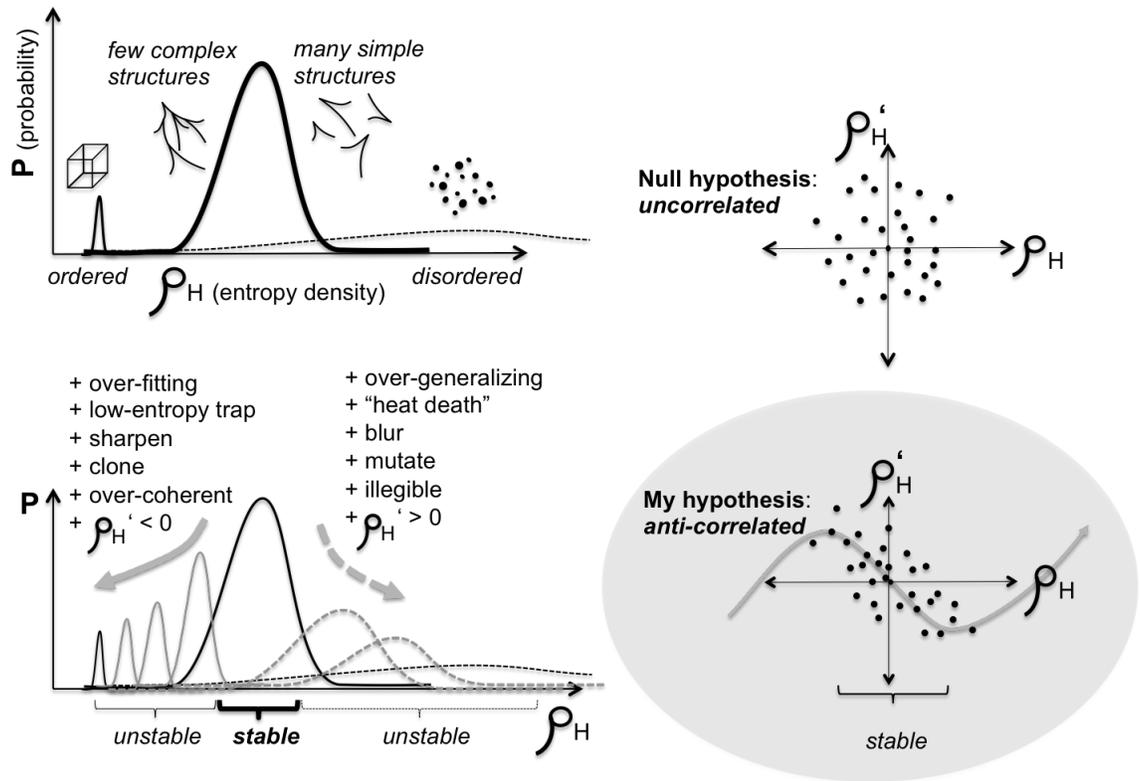

I hypothesize that population-distribution stability exists in a zone near zero entropy-change, bordered by an annulus of runaway instability on either side.

On the extreme right, where entropy increases, lies death by diffusion, in which the population becomes ever-more-diverse as its self-replication standards become weak.

On the extreme left is death-by-duplication, as over-strict enforcement of old standards and a lack of new diversity leads to ever-narrower distributions which span ever-tinier slivers of parameter-space.

Such over-focus, like over-fitting, leads to sharpening traps whose low entropy is no longer robust to random events, such as environmental fluctuations. Stability only is possible in between, where the ensemble is refined enough to persist, yet diverse enough to resist. Adding diversity is like adding uncertainty: it blurs expectations, makes them less specific in space and time. But that blurring has to happen in a specific way: in that central stable region, the distribution must have consistent inhomogeneities such that ensembles with marginally higher entropy cause the overall entropy to *decrease*, and vice versa.

With this condition, population distributions propagate across time with the artificial narrowness of solitons, because the passive blurring forces of diffusion and dispersion are themselves actively balanced by active sharpening forces of amplification and selection.

*Questions:*
Do chain letters and viral media follow this rule?
Must specialization fight symmetry?
What should I do when I don't know what to do?
How can I manage a team so we don't get stuck?

## H2.0 Stabilization and homeostasis are fragile in multiple ways

**If the self-regulation of an agent or population is stable in some environment via parameterized instincts, then outside of that parameter-space the agent/population will likely encounter control problems at multiple timescales, which at best impair its function, and at worst make it actively destroy itself and its compatriots.**

\* \* \* \* \*

Active stabilization (*homeostatis*) is a feedback system like a car's cruise control, set to keep one thing steady (like speed) in spite of "hills and headwinds." To ensure the circuit stays stable, its design specifications regarding environmental concentrations and correlations must be respected as an inviolate statistical contract. Outside a control system's original parameter space, the contract is null and void. Control won't work.

The contract is to keep a small number of parameters stable in a small region of space and time, by exporting low-entropy reside or "waste" influences to later times and other spaces.

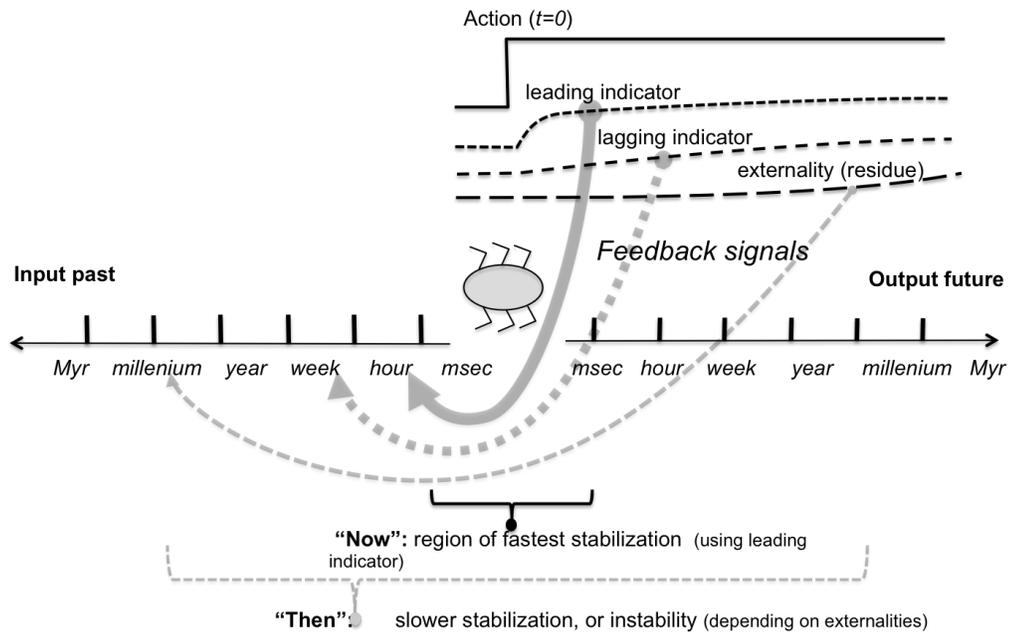

I hypothesize that there are four general mechanisms spanning all temporal scales which may render otherwise stable control systems non-functional. All of these mechanisms ultimately involve over-sharpening or "over-fitting" dynamics, in which old data and strategies prevent adaptation to faster-changing circumstances.

1) **The slow accumulation of waste**. Residue or waste is any enduring byproduct or externality, however simple or complex (usually if not always low-entropy, as the result of selection) . Hundreds of millions of years ago, life-forms exhaled a reactive by-product, oxygen; oxygen ultimately changed the biosphere and killed off many of the organisms which created it. Polluting the environment with waste of any kind is the slowest and simplest way for an agent to hurt itself.

 2) **Changes in in homeostatically-relevant parameters** of the environment. Even a single-dimensional system controlling a single variable can be tricked into instability if parameters of the environment change: signal concentration, visibility, gradients, enrichments, distribution, availability, timing, or variability in timing.  The situation becomes more complex when multiple variables, timescales, and cross-interactions are involved; the more possible feedback pathways, the more likely one of them will lead to instabilities.

**3) Leading Indicator Dependency**. A self-regulating system succeeds when, in its environment, the system's short-term appetite (*leading indicator*) is positively correlated with fulfilling long-term needs; but if the environment flips that correlation, the system enters a temporal sharpening trap 'addicted' to what makes the problem worse, and becomes stuck.

**4) "Explosive" residue.** A hypothetical but presumably rare situation is one in which a specific kind of residue accumulates slowly, but then ultimately produces fast effects because some reaction scales super-linearly with residue density, with a critical threshold. Because such super-critical behavior would appear almost instantly (on evolutionary scales), adaptation occurring would seem difficult. The most explosive of such super-criticalities would be like stampeding animals, as agents trigger one another via fast visual and acoustic communications. (See H5.0)

*Questions:*

Do we crave sugar for sweetness, or for speed?
How can I curb my own appetites?
Does online socializing make you feel lonelier?
Does Life produce entropic residue along with chemical waste?
Do market forces make consumer-product quality decline?

## H3.0 Representing spacetime requires micro-timing and mega-assumptions

**Any optimally efficient real-time simulator/controller must assume everything it can, autonomously steer its sensors, and squeeze out all the timing information possible from its inputs, whether a brain modeling muscles from distributed mechanoreceptors or an autonomous vehicle modeling its environment from distributed sensors.**

\* \* \* \* \*

With limited data, a real-time spatial-imaging (*tomography*) system must face algorithmic constraints and tradeoffs, such as those confronting self-driving cars. Those systems must infer continuous-valued shapes from discrete data sampled at discrete times. The spatial and temporal resolution of such simulation and control systems are known to depend directly on the quality of the data, and on the precision and robustness of the statistical assumptions used to process it.

I hypothesize that brains can be understood alongside such real-time tomography systems, and face the same constraints:

1) **Microtiming matters.** The precision in space (*mm*) and in time (*msec*) of the moving map depend directly on the autonomy (*output entropy*) and timing precision (*latency jitter*) of the round-trip path from actuator back to sensor. To move with precision, the system must be very sensitive to sudden input changes.

2) **Focused priors matter.** For reasons of speed and stability, as many priors as possible should be wired-in rather than learned. The more narrowly-focused the space a simulator has to search, the more resolution it can wring from data in that space. A nearly ideal simulator will in advance (e.g. in hardware) assume the continuous three-dimensionality of its target space, the constancy of objects and motion (e.g. momentum), the continuity of surfaces, the scaling of distance-change with velocity, and so on. In particular, a small target space means a large null-space, whose presence systematically **cloaks** actual inconsistencies in the model, in the same way the brain cloaks the retina's blind-spot from perception.

3) **Autonomous appetites**. A system must have the capacity to steer its sensor of its own accord, unscripted by outside influences, in order to simulate a rich environment with limited sensor information. The system's informational appetites—the locations and types of input it desires and avoids—will vary moment-by-moment according to both short- and long-term calibration needs. In general, the more decalibrated the system, the more it seeks out quickly-available, coherence-enriched inputs.

*Questions:*

Why do we feel emotions when we recognize places or faces?
Do we feed our algorithms better training data than we feed our brains?
Why does VR make you sick?
How did Tesla's "MECHANICAL THERAPY" machine work?
Could a robot feel its body with vibrations?
How precisely can we feel our bodies?
Do our nervous systems use microseconds?

## H4.0 When in doubt, ping!

**An active sensorimotor system must interact with its environment like a submarine might, arranging its output activity along a spectrum of increasing coherence, starting with passive listening and tracking (*silent running*) at one end to active coherent illumination (*pinging*) at the other, with the costs of pinging reserved for reducing uncertainty.**

* * * * *

To keep itself in calibration, an active models may choose the patterns it emits. Because outputs consume energy and reveal oneself to the world, smaller and subtler outputs are preferable, meaning that outputs should be reserved for when they are in fact needed to improve the model.

Pinging is one such recalibration strategy. If optimized correctly, pings create a loud enough return signal to re-establish timing lock using only a few pulses. However, pings cost not only energy, they require fixing the reference frame for the duration of the ping and echo (thereby losing continuity and resonant connection), they fracture continuous expression, and they create distracting external signals which may draw unwanted attention.

I hypothesize that as a general rule, a good strategy for a microtiming circuit to keep itself in calibration and its model trust-worthy is to invest energy in coherence-power whenever internal timing precision declines or uncertainty increases. A self-calibrating system ought to as a matter of principle create ever-more-coherent and high-amplitude outputs (*pinging* or *flailing*), the worse its sense of awareness and control becomes.

***Questions:***

Why is the world so stressful?
Are some people natural "pingers"?
Are some people natural listeners?

## H5.0  Mediated communication becomes infected with pinging

**While pinging reduces uncertainty for the lone individual sending the ping, that extra coherent energy appears to others as noise, increasing their uncertainty (as an *externality*); to mange that tradeoff, some optimal uncertainty-to-ping ratio must be wired-into the informational instincts of individuals social beings sharing quiet, organic environments, instincts which unfortunately become counterproductive when they communicate with amplifying media, whose gain creates a chain-reaction dynamic (*microphone feedback, amplified echo-chamber, behavioral epidemic, Fabry-Perot laser, tragedy of the commons*).**

$$*  \qquad *  \qquad *  \qquad *  \qquad *$$

Let a property of a communications medium be *transparency*. "Transparent" means the media doesn't absorb or suppress any particular kinds of message (*neutral*), it doesn't amplify or promote messages on its own (*stable*), and it transmits the same in all directions (*symmetric*).  A transparent medium does not impose its agenda on those using it.  In that sense air and skin are transparent; symbolic media like books and screens are not.

I hypothesize that when humans use non-transparent media for communication—in particular low-bandwidth, amplifying, error-prone media like digital media—our natural human instincts to ping one another for reassurance will accelerate catastrophically and endanger collective mental health.

*Questions:*

Why aren't phone calls as fun as they used to be?
Can media be toxic?
Is pinging an unnatural strategy for nervous systems?

## H6.0 If scaling and incentives are the problems, then entropy and affection are the solutions

**If human nervous systems have been increasingly decalibrated by a low-entropy material system incentivized to make us even more so, a workable solution must reverse all those effects at once, i.e. establish local, temporary human-safe spaces (*trust-enhancing*, *sacred,* or *embodied containers*) where people may experience minimal complexity and interference, and maximal sensory entropy and physical affection, in order to collectively promote and recalibrate the human birthrights of goodwill and resonance.**

\* \* \* \* \*

The human nervous system, like all nervous systems, evolved to process continuous natural statistics while navigating through continuous natural environments. Now the majority our sensory inputs are selected if not outright synthesized, and our choices are ever-more digitized and constrained. This reduced sensory entropy and increased choice-dimensionality make sensory processing harder, and lead easily to fractured world-models and motor maps, whose fractures are hidden from consciousness by cloaking. The fracturing epidemic is caused in part by buildings, lighting, furniture, rules, and habits which are bad for (i.e. decalibrating to) our sensory and social systems.

I hypothesize that human experience and function will improve in deliberately sculpted environments  (*containers*) designed to feed human socio-sensory needs according to universal and neutral principles of symmetry, continuity, and entropy.

1) **Symmetry** says all humans in the container are equal in interaction, so incentives are forbidden.
2) **Continuity** says smooth space and sentiment, not hard walls nor rules, be used to separate, so no sharp lines in space, time, nor judgment.
3) **Entropy** says variety is precious in facial expressions, vocalizations, and styles of motion, but it also says proximity, visibility, and touch provide positive bandwidth benefits.

*Questions:*

How can I host a productive discussion?
What is the optimal way to negotiate?
What is the best way to dance?
Are economies of attention and of behavior moral hazards in and of themselves?

## H7.0 "Paleo everything but violence" provides sensorimotor nutrition

**If the "paleo" human nervous system evolved to collaborate with close kin in a physically simple space filled with organic things, then interaction with equivalent sensorimotor patterns at various scales, whether smelling a rose, planting a garden, or climbing a tree, ought to at least recalibrate a nervous system's dynamic range, if not outright heal its acquired fractures.**

\* \* \* \* \*

If a paleo digestive system and its appetites evolved to process naturally available organic food, then a paleo nervous system and its appetites would have evolved to process naturally available organic inputs. That natural variability applies to three aspects of sensory experience: 1) to the extremes of sensation experienced, 2) to the continuity of experience, and 3) to the entropy of the inputs. Unfortunately, those same natural human appetites make us avoid sensory extremes (say of cold and hot), make us notice and seek discontinuities, and make us choose unambiguous inputs over ambiguous ones. In short, the fracturing of our nervous systems is caused by the presence of too many man-made information sources and restrictions, combined with innate appetites choosing them over natural alternatives.

I hypothesize that a simple cure is reversing those statistics. That is, to increase one's sensory range, to reduce the dimensional entropy of one's interactions, and to re-introduce slowly-changing organic (e.g. boring) things into one's environment.

How does it help me to visit a forest
Should I be thankful for discomfort?
Does A/B testing damage learning "object-constancy" in children?
Is oculocentrism a natural orientation?

## H8.0 Humans evolved to resonate ecstatically

**With humans as with bosons, there must have evolved an attractive force to make us congregate: the force of taking potentially unlimited pleasure in affectionate companionship.**

For small kin-groups foraging nomadically, the vast majority of social interactions would have been with friends, not enemies, so human sociability makes sense from basic principles. Furthermore, the same evolutionary logic which justifies ecstatic pair-bond pleasures between lovers, or between mother and infant, would also apply to multi-person interactions.

I hypothesize that humans have a nascent capacity for intense, monomanic social pleasure, which might be dubbed *social ecstasy*, which is every bit as strong as our capacities for chemical, physical, or sensory pleasure. This pleasure would be based in the same vibration-management circuits which already drive spinal alignment and dyad resonance (see Appendix B for a possible mechanism), and would bond families and friends as tightly as lovers. Practice would promote nervous-system recalibration and healing.

*Questions:*

How can I resonate?
How can we make love more contagious than fear?
Can mis-tuned instruments tune each other?
Can people heal each other with ultrasonic coupling?
Can human skin feel delicious?
What would an ideal Yoga or dance studio look like?
Can group vibration heal us?

## H9.0 Symmetric spinal health syndrome

**The documented benefits of practices like Yoga, Pilates, Feldenkrais, aerial dance, Tai Chi, Ecstatic Dance, Capoeira, and other movement/meditation practices share a common cause in the spinal transformation they induce, because those practices provide**

**optimal 4-D sensorimotor input and intention profiles to recalibrate and anneal the body's proprioceptive midline map, and thereby to optimize real-time sensory sensitivity and fluidity.**

* * * *

Spinal ecstasy is Nature's solution to a thermodynamic problem: there are many configurations which curl and kink a spine, but only one which straightens it. In other words, to reach that low-entropy but essential condition of straightness, a spine needs to *want* to be straight. It needs a built-in entropic pleasure to urge it toward the single best configuration (See Appendix C)

I hypothesize that resonant ecstatic states exist: hyper-focused, hyper-simple, pleasurable, healing, and so all-absorbing they resist conscious recollection. Because any rod vibrates the fastest when perfectly straight, the mechanism for ecstatic spinal straightening would be an in-built drive for the spine to resonate at the highest (ultrasonic) frequencies possible, which occurs when local proprioceptive ambiguities have been smoothed out.

*Questions:*

Why does my body become achier and stiffer as I age?
How does my mind work?
Do data problems cause our itches, aches, & pains?
Why do people love activities like Yoga so much?
How can I supercharge my (Yoga/Pilates/dance/pole) practice, or at least feel younger?
Does central spinal activation make me feel better?
Do vibrational therapies share common mechanisms?
Can I become a hyper-athlete?
Why do my joints pop?
Are the muscles I feel inside me impossible hallucinations?
 Why do people benefit from aerial sports?
Should my body feel detailed, or simple?
Is the "chakra system" a consequence of spinal software architecture?
Does music ring our spines like violins?
Can ultrasonic senses be trained?
Can you hear the shape of a drum, or of a cave?

## H9.5 Helping the irresistible force beat the immovable object

This is not a hypothesis, but a call to action, so it must be accepted *before* it can be tested. That makes it only worth a half-point, but that's better than nothing.

**The immovable object is E.M. Forster's *Machine*, the collective sharpening forces of material culture which pry people apart; the irresistible force is the collective computational power of ultrasonic human resonance, growing in synchrony over ever-larger scales of space and time; the balance can be tilted if people remember that the resonance of human love has a vastly higher bandwidth and pro-human orientation than does steady material accumulation, and that optimism is in fact the optimal resource-allocation strategy.**

\*     \*     \*     \*

On the one hand, the view seems grim. Every technology and interaction-style plaguing us humans, from physical padding to digital pokes, is the nearly inevitable, incremental multiplication of two principles: the principle that brains like occasional discontinuities and moments of recognition, multiplied by the principle that we make things that we like. We are the only creature perceptive enough to fall for patterns of pixels, and the only creature deft enough to make them, so of course we do. Those behaviors feed on each other, until our nervous systems are dazzled and distracted. That human condition is a continuous extension of the inexorable entropy-reduction in Earth's biosphere. On that grand tragedy, even "capitalism" and "material culture" are but small-time villains.

On the other hand, humans evolved for resonant love in small groups, and with safe guidelines can easily rediscover it. That ultrasonic resonance surpasses thousands-fold the computational speed of "thinking" as we think of it, and as a bonus runs on our native processing architecture. Plus, it feels good, heals you, heals them, and helps everyone like each other. What's not to like? With a straightforward understanding of how our brain+body systems work and what we need, people everywhere will rediscover our own spines, and reconnect with our family, friends, and loved ones. Each resonant connection will enable subtle (but not auditable!) micro-interactions which facilitate success—on *average* but not guranteed—among that tribe, and facilitate resonance with those outside it. Even just two people resonating together over time, like spouses or collaborators, can produce amazing things, and that effect scales up to the dozens we were born to interact with.

The good news is, it doesn't stop there. Our nervous systems evolved for super-critical resonance in groups of just a few. Now we have millions at a time. In principle humankind will automatically self-organize if we come together in the right ways.

The wrong ways are with organizational sponsorship. Every bit of funding, branding, advertising, national or corporate propaganda, uniforms, or participation constraints sculpts the entropy of the otherwise-neutral resonance-cloud, which physicists might call a *boson condensate*, and breaks the natural human symmetry of one warm body near others.

The right ways are with pre-existing resonant groups who experience social happiness already, and want to help spread more of it: yogis and dancers, matriarchs and patriarchs, happy couples, groups of friends, close co-workers, long-time teams, family businesses, small-town gossips, community organizers, local politicians, and in general those with the highest ratios of social to monetary capital. They can be the human seed-crystals, while the abstract Framework can concentrate the rest of us into cooled molten sugar, ready to instantly crystalize.

"The rest of us" means those with means and training: serious intellectuals, philanthropists, community organizers, church leaders, engineers, scientists, computer scientists, chieftans, and emperors. Our job, once we reconnect with our common resonant heritage, is to tackle the problems of the ages all together, all at once.

Each skill is necessary. Among abstract thinkers, theoretical physicists and data scientists can build hypothesis-validation software which makes plotting truth as straightforward as plotting data. Business leaders who want to steer the world from profit-only toward profit-plus-humanity can invent new business models which might do so. Attorneys and judges can figure out how to morph a discontinuous, precedent-driven legal system into a continuous, principle-driven one. Economists can find a model for economies which doesn't put trust up for sale. Dancers can remind us how to move. Bodyworkers can heal and teach embodiment. Mothers can be attentive to their kids, and make them be attentive back.

The grim view is that Life on Earth is doomed to die of entropy-reduction. The happy view is that ten billion human beings are alive and still can love people close to them. The necessary hope is that when humans get a chance to act on Love, we do it well.

## Acknowledgments

This work is part of a larger, co-equal collaboration with Criscillia Benford which includes the informational structure and eigenmodes of narrative.

Not everyone can understand these hypotheses or make use of them. I believe the people below still living could do so, and I would be personally flattered if they did.

*Friends and family whose conversations have influenced me:*

Ed Softky, Wendy Dando, Minet Sepulveda, Anne and Eric Doehne, Kwabena Boahen, Adam Safron, David Galbraith, Matt Ogburn, Mark Griffith, Peter Hollingsworth, Mitch Ryan, Tristan Harris, Sheldon and Marion Softky, Michal Peri and Tyler Holcomb, Allison Hudson, Michele Fabrega

*People whose personal interactions have influenced me:*

Freeman Dyson, John Hopfield, Carver Mead, Kip Thorne, Stearl Phinney, Christof Koch, Bruno Olshausen, Tony Bell, Douglas Rushkoff, Kwabena Boahen, Josh Golin, Jeff Hawkins, Jerry Gollub, Bill Newsome, Sebastian Seung, Bill Bialek, John Rinzel, X.J. Wang, Zhaoping Li, Tobi Delbruck, Dan Kammen, Dileep George, Wyeth Bair, Dan Murnick, David Van Essen, Peter Hart, Severo Ornstein, Terry Sejnowski, Lloyd Watts

*People I only know by reputation:*

Elon Musk, Max Levchin, George Soros, Bill Gates, Mark Zuckerberg, Warren Buffet, the Dalai Lama, Noam Chomsky, Barack & Michelle Obama, Nate Silver, Randall Munroe

## Appendix A. Warrants: The Geometric Laws of Nature

In the Beginning was the Singularity. Ever since, Entropy has been increasing throughout the Universe.

Except here on Earth.

Entropy is "information" by another name, based on the same equations for counting combinations. But entropy/information is not the deepest concept in the Universe. Spacetime is.

If one views the universe in data terms, spacetime is the deepest architectural layer. According to Hawking, Hartle, *et al,* spacetime has *no boundary condition,* meaning it is continuous in every possible sense. The universe is *multi-scale,* meaning it has no particular size. Structurally, it has no preferred center, it just is. It is *isotropic,* meaning it has no preferred direction. The "shape" of spacetime embodies fundamental *symmetries* such as translation, rotation, and dilation; it contains implicit *reference frames* such as spherical, cylindrical, and Cartesia; and it allows fundamental shapes, such as the *Platonic solids*. Spacetime can twist, shear, and dilate, but not break. By software inheritance properties, any substructure embedded in spacetime can therefore also twist, bend, or expand.

Spacetime has no fractures; it is seamless, continuous, and connected.  To add "...even as time evolves" makes the statement redundant. Spacetime (and of course the things in it) follow the continuous Laws discovered by Newton, Maxwell, Einstein, Feynman *et al*. In software terms, spacetime follows a *continuous interface format*.

But following real-valued rules does *not* necessarily mean spacetime acts like real numbers in other ways.  In particular, as Heisenberg *et al* pointed out a century ago, spacetime does not have infinite fine-ness (*resolution*) and thus infinite information. It is blurred and uncertain. There is only so much information in any zone. According to Wheeler, Kantor *et al*, information might yet be more fundamental than spacetime itself.

Discrete, distinct objects from particles on up fracture the elegant simplicity of spacetime. Now separate things can be combined in exponentially many ways (*combinatorially*), allowing complex structures like Life.  *Entropy* appears (Clausius *et al*), and makes things complicated. Fortunately, simplicity reappears at the upper bound of entropy, where continuous, low-dimensional Laws like temperature and pressure once again compress the motion of infinitudes of molecules into a couple scalar parameters.  The continuous Laws of Entropy (including Shannon's Laws of Information) do not describe fundamental spacetime structures, but aggregates of known things. By the same token we understand those math-made laws even better, and can intervene in their details.

Science is a form of description. Occam, Bayes, Shannon, Mackay *et al* showed that the most efficient descriptions are simple.  So the continuous Laws of Nature form the most simple, low-entropy descriptions possible outside direct sensory experience, and thus the best ones.

The principle is that principles trump evidence, and the evidence agrees. Principles are always involved in gathering data, in interpreting it, and in planning action. Principles are what unites "What is?" with "How then shall we live?"  Most crucially, "principles" scale differently than "data."  Big data usually beats small data, but the simplest Root Principles are always best.

Besides Spacetime itself, the most principled of the Elementary Principles is that of Entropy, sometimes known as Shannon Information.  Entropy decreases in closed systems, but entropy can decrease locally when provided with energy sources and entropy sinks (*non-equilibrium thermodynamics*). Such a situation allows for the first core principle of Life, *active stabilization*.

On the one hand, an active stabilization strategy can be described as *homeostatic*, that is a single variable balanced between opposing forces, whose tug-of-war create zones of (meta)stable self-regulation over time, as the single parameter tracks inputs from outside. The same system might also be described through its *resource allocation strategy,* i.e. how does the system distribute its resources (energy, time, location) across a given parameter space…blurred and broadly distributed, or specialized, narrow and focused (Hills)? This approach analyzes the whole distribution, not just a central value, and can thus applied to populations.

Stabilization is only possible using *sharpening* forces. Sharpening is the most useful umbrella term for the host of entropy-reducing mechanisms of life: self-replication, edge-enhancement, selection, amplification, specialization, etc. Unlike those, sharpening describes not a specific physical mechanism, but the abstract transformation applied to any probability-distribution function, so it can be used to describe all possible entropic consequences.

The universality of sharpening is central: once started, sharpening as a process may be unstoppable, since there seems to be no threshold of size, mass, or complexity above which the process would reverse.

Geometrically, the algorithmic ease of active stabilization depends on the (low) entropy of the parameter space being controlled and the (high) entropy of the patterns in it. The data-constraints of stabilized systems involve the same principles as do scientific theories (*Occam's Razor*) and algorithmic training (*statistical inference*). As a general rule, low-dimensional or low entropy spaces are easier to search than high-dimensional or complex ones (the *Curse of Dimensionality*). But for the data *within* those spaces, high-entropy, variable, multi-resolution data (e.g. natural or naturalistic inputs) provides better long-term training than do low-entropy "test-patterns," which create fractured models due to over-fitting.

These constraints have always been at play as evolution told life how to reproduce, and told our brains how to operate our spines and bodies. But now we as humans receive our instruction not from Nature but from material society, so we are not told everything we need to know, even though the same constraints apply. Thus, understanding how the Warrants above apply to generic Problems of Life and Mind (Lewes) ought to provide helpful, neutral, common guidance for humans and those who support us.

## Appendix B. Detailed questions

## H1.0 Stably evolving distributions must balance themselves between narrowing and broadening

> **Do chain letters and viral media follow this rule?** The entropy of a text (in bytes) is the number of characters. So any population of "viral" texts which persists has an entropy density and a mutation rate. What are those amounts? Do they show the inverse correlation hypothesized here, i.e. short letters mutate more, complex ones less?

**Must specialization fight symmetry?** In most people one hand is dominant, left or right, one leg, and one eye. Those asymmetries are functional insults to a bilaterally-symmetric body, whose motor-architecture overall would be simpler (and thus better) with no asymmetries at all. This predicament reflects the following paradox. On the one hand, the most motor-friendly resource-allocation strategies are simple, uniform, and symmetric, spreading both motor activity and the brain's representational space evenly. Specialization damages that equipartition by focusing on only a few postures or activities. Is this an example of the focus-vs-blur tension? If so, is there a self-regulating mechanism, or will specialization (e.g. in careers, genres, and ultimately life-forms) increase without limit?

**What should I do when I don't know what to do?** The instinct for more certainty is perfectly natural, built into native informational appetites, but it only works effortlessly in the wild. In civilization, humans must make an effort not to make efforts. In this Framework, managing uncertainty is central to life itself, because life forms are data omnivores, always hungry for more. So for reasons that can be spun out at length (if you like mathematical ideas), the advice is almost always the same, regardless of the specific situation. Call it the "universal solution to sharpening loops and traps." If the most generic problem of Life is over-focusing, then the solution is to un-focus. Don't ask for quick results or instant gratification. Don't interrupt. Don't micro-manage, or even manage. Don't audit, worry, or self-recriminate. Don't make decisions. Trying to record or control such a situation makes it worse, like a drowning person flailing. As in the stock market, waiting and accepting are the most successful algorithms on average, and the best for everyone else.

**How can I manage a team so we don't get stuck?** As a species, humans spent a couple million years in small family groups, probably the tightest teams to ever roam the Earth. So cooperation is in our DNA, and it ought to be instinctive. And it was instinctive, everyone tracking everyone else, as long as we were roaming real savannahs in search of real food. Unfortunately, modern environments are more abstract, and we don't know, hear, or see each other as clearly as our instincts expect. So the instinctive, sub-conscious information channels which ought to make us sense when we're getting stuck don't work, while the big, fat, loud, obvious channels like metrics and emails and quotas do work. Those clunky channels are unambiguous. Relying on those channels is the problem, because of all the simplifications and categorical approximations those channels impose. Those channels quash the diversity our nervous systems need. So in modern environments, the single most important thing for managers to do to combat the structural tilt toward standardized groupthink is to keep diversity alive in every form: neuro-diversity, acoustic diversity, cultural diversity, activity diversity, goal diversity. A good implementation will not be based on rules or incentives, but based on liking other people and wanting them to like you back.

## H2.0 Stabilization and homeostasis are fragile in multiple ways

**Do we crave sugar for sweetness, or for speed?** In evolutionary times sweet taste was well-correlated with nutrition content. In that environment sweetness would be a leading indicator for nutrition as a whole, because one could taste it immediately (unlike, say, vitamins, whose benefits take time to feel).  How much of a substance's addictiveness comes from the *informational high* of quick information, versus the quantity of the drug?

**How can I curb my own appetites?** Seeking subtlety brings us back to bandwidth. The tragedy of the human condition is that we have to worry about curbing our appetites at all. Animals were born with appetites so we could chase them, not resist them. Back then, in Nature, getting too much wasn't usually a problem. But now it is, not just with sugar but with news from friends.  Now instant gratification is always moments away, and it turns out too much is bad for us. So we—this current generation—is facing more temptation, and more technologically sophisticated temptation, than any generation in human history, and we haven't any training to resist it.  So we have to be kind to ourselves. No organism ever evolved to resist what it wants.

While we can't make appetites go away, we can choose to focus on some aspects over others. In general, the simplest, best strategy is to aim for subtlety over obviousness: aroma over sweetness, vocal texture over words, breath-sound over vocal texture.

**Does online socializing make you feel lonelier?** Do the informational hits from online communication present themselves as actual social interaction? If so, does their quickness and quantifiability present itself as a leading indicator of the real thing?  If such quick interactions fail to deliver the long-term bandwidth and satisfaction the nervous system expects, i.e. if they serve as *false leading indicators*, will a decalibrated nervous system still seek yet more of them?  As such influences make social beings progressively more alienated, will they crave those leading indicators even more desperately?

**Does Life produce entropic residue along with chemical waste?** The many small, simple creatures which long ago first created the oxygen externality left as their residue a very simple waste product, a diatomic molecule. Later, more complex lichens produced a more complex waste chemical, *calcium oxalate.*  Macroscopic (low entropy) structures left by pre-human creatures are mounds (*stromatelites*), coral reefs, termite mounds, and bowerbird bowers. The structures left by humans dwarf even these, being as large as nation-states and as enduring as concrete. Is it the case that the enduring physical structures created by life have ever-increasing individual complexity, yet make the world as a whole less diverse?

**Do market forces make consumer-product quality decline?** Can a consumer determine the price of an item more quickly than its quality? More accurately? Is the consumer compelled to use the faster and clearer source of information over the slower and more uncertain one? Would such a purchasing strategy lead to consumption of cheaper products whose quality decrement cannot be determined at purchase time? Would such consumption lead, in aggregate, to products of lower quality?

## H3.0 Representing spacetime requires micro-timing and mega-assumptions

**Why do we feel emotions when we recognize places or faces?** Many animals get excited to see familiar people or places, but none so much as us, in great part because our perceptual systems are so sophisticated. *Homo sapiens* could see the outline of a bison in the smudges of charcoal on a cave wall, and now we see far-away lands in the flickering pixels on our screens. The way Nature designed it, we not only *can* recognize stored images better than other species, we *want* to recognize them, as if the new ability came with an appetite for using it. Combined with the thrill or jolt of recognition also comes whatever other emotions tagged that image when it was stored, revived as echos.

**Do we feed our algorithms better training data than we feed our brains?** An algorithm is only as good as its training data, which ought to be neutral, naturalistic, and unbiased. The algorithms which guide our cars and computers receive such high-quality training data. Many patterns consumed by humans are different, because they are generated by machines in order to measure us or influence us. In data terms, the biases, quantization errors, autonomy constraints, and especially microtiming damage contained in such input profiles would be unacceptable to any self-respecting self-learning algorithm.

**Why does VR make you sick?** So-called "simulator sickness" has been around for decades. In retrospect, it seems obvious that any sensorimotor environment which does convince your eyes that you're moving, but **doesn't** convince your body, is doomed to confuse the nervous system in a very deep way. Part of the problem is that VR can never give you the microtime signals you need, because the VR system has to wait for your eyeballs to move, while your brain already knows how they'll move before they even start. The deeper problem, however, is strategic: it's very bad data-hygiene to make your visual system compete with your mechanosensory system, because the two evolved to agree. While the data-conflict is more superficial with less invasive sensory stimulation, it exists even with smartphones. We can become socially sick when our mediated social inputs become as important to us as real-life ones.

**How did Tesla's "MECHANICAL THERAPY" machine work?** Nikola Tesla used a high-powered vertical ultrasonic vertical oscillator to create an "artificial earthquake" which shook buildings in New York city, and on humans induced visceral pleasure and digestive benefits (he called it MECHANICAL THERAPY; Tesla's friend Mark Twain loved it). How might it have worked? A classical mechanics analysis of a vibrating object would order the body's eigenmodes in the space domain, largest amplitudes first. But for information transmission (bandwidth and resolution), the frequency domain dominates, ranked by highest frequencies first. In an upright, compressive structure like a building, or a body, the highest-frequency eigenmodes are vertical and longitudinal. A body in the frequency domain poses unorthodox questions: What is the highest-frequency vibration possible in a human spine under optimal organic conditions? What are "optimal conditions" in both anatomical and attitudinal terms? How close are optimal conditions to what reigns in modern bodies? How collimated are those vibrations in myofascial tissue? What is the upper bound on mechanical information flow (via Nyquist) from such signals? How does that flow relate to flows from other sensory and "cognitive" sources? How much active amplification and sharpening is involved in sustaining those vibrations? Are there harmonic patterns in spinal vibrations, as there are from linear musical sources? Would such patterns correspond better to eigenmodes, to attractor states, or to sculpted, synchronized wave-packets?

**Could a robot feel its body with vibrations?** Can a robotic control-system be designed which infers its self-model from internal vibrations rather than from purpose-built position and acceleration sensors? Would active sharpening of those vibrations help that process? Would the process work better with continuous-wave carriers whose frequency changes, or with sculpted solitons whose timing changes? What mechanical relationships between activator, structural member, and joint-hinge (e.g. impedence-matched density and stiffness) allow the most precise vibration-sensing? Are those mechanical and vibration-management principles evident in human bodies?

**How precisely can we feel our bodies?** If by "proprioception" we mean the sense of body shape, motion, and forces it feels via mechanoreceptor channels, what is the microtime jitter of such inputs at their source, individually and collectively? How does it vary with temperature? How does it vary with sensory attention? What is the lower frequency bound of mechanical (acoustic) signals capable of entraining such mechanoreceptor precision? What is the upper frequency bound of mechanical signals that tissue can sustain passively? Actively? What are the algorithmic constraints on inferring proprioceptive sensations from such signals in the ideal case? How close do our brains come to that ideal?

**Do our nervous systems use microseconds?** How small are temperature fluctuations at the second scale in warm-blooded spines? How many microseconds of variability (*microtime jitter*) do those temperature variations induce in action-potential roundtrip transit time? How great is microtime jitter originating from all distal (tail) vs. cervical spinal sources? What is the corresponding microtime jitter entirely within temperature-regulated cortex?

## H4.0 When in doubt, ping!

**Why is the world so stressful?** Paradoxically, life feels bleak where markets and technology work their best. Markets are amazing at providing what people will buy, the quicker the better. Technology is amazing at giving markets what they want. In combination, those two forces provide two effects which are fine in small doses, yet toxic in large doses: appetizing products, and persuasive ads. Product-designers have done their jobs so well we can't resist those gadgets; advertisers have done their jobs so well we don't know why we do their bidding. The end result of that success is a market built on hailing our "inner pingers" via instant-gratification opportunities. The cleverest of those markets, like social media, have monetized our urge to ping each other….all they have to do is provoke the pinging and collect the tax. An especially potent way to prod people into pinging is via remote interruptions. Unfortunately, interruptions are bad for our nervous systems, and cause enormous stress to us.

**Are some people natural "pingers"?** Pinging as a practice might lie deep in the nervous system, making pinging a life-strategy and not just a one-off choice. To be a natural pinger, someone would prefer sudden transitions as a matter of principle. Such a person might self-stimulate by fidgeting, cracking their knuckles, or biting the fingernails. In conversation, they might interrupt, ask lots of questions, and dominate the reference frame. In life, they might make sudden changes of career or direction. Such behavior might be called "attention-getting," and might be associated with deficits in social trust and emotional sensitivity. Are pingers' spines stiffer than normal?

**Are some people natural listeners?** Do people listen more when pings might cause problems, as in crowded classrooms or interaction with higher-status people? Do typical listeners also specialize in other forms of highly sensitive, long-duration data acquisition? Does listening involve more physical and emotional resonances than pinging? Do listeners go more by "feel" than rules? Are their spines more supple (H9)?

## H5.0  Mediated communication becomes infected with pinging

**Why aren't phone calls as fun as they used to be?** When telephones were first deployed across America, the audio quality was so good that musicians used to practice together using the shared neighborhood telephone line, then called a "party line." That practice was so popular it had to be prohibited. Even thirty years ago, land-line telephone calls could still carry whispers and subtle tones of voice, so that happy conversations could go on for hours. That remarkable vocal reciprocity was possible because fixed wires were in place, making bandwidth basically free. But with mobile calls bandwidth isn't free, so carriers are under relentless financial pressure to reduce audio quality to just above the point where people quit. Furthermore, the algorithms involved make the resulting "compression artifacts" such as gurgling, glitchiness, and dropouts not only common, but difficult for a nervous system to anticipate or correct for. In essence, modern wireless-voice communication has squeezed out the microtime meta-data our nervous systems need to perform calibration and trust, leaving only recognizable "content."

**Can media be toxic?**  On average, the internet makes people look stupid, heartless, and threatening, but only because it isn't made of air.  Here's why. A physical medium (air, water, wire, radio, smoke) transmits signals. Humans evolved only to communicate through transparent, instant media like air, not through any medium which picks and chooses what to send, which inserts delays, or which keeps the message around longer than a second or so. Does a non-transparent medium convey trust in physical things as well as the things themselves do directly in proximity? Are its informational properties like bandwidth and latency correlated with the trust it conveys? Does real-time interactivity improve that trust? Does better microtiming jitter improve that trust? Do these effects apply for trust between people? Does one-way digital broadcast form a significant part of human informational diets? Does such broadcast transmit fear, uncertainty, and doubt more efficiently than hope, assurance, and love?

**Is pinging an unnatural strategy for nervous systems**?  Is pinging *as a strategy*  unstable?  Is it topologically prone to feedback-traps that listen-only strategies would not fall into?  Do nervous systems thrive on truly continuous-time strategies, in which the system's reference frame co-evolves as data steadily accumulates? Should listening be the default mode for any nervous system?

## H6.0 If scaling and incentives are the problems, then entropy and affection are the solutions

**How can I host a productive discussion?** The best discussions are those in which the participants are connected symmetrically in as many ways as possible, with as high interpersonal bandwidth as possible. So for purposes of illustration, an "ideal" discussion would have these features: 1) Everyone in the same quiet, visually appealing (but not distracting), non-flickering environment, to preserve microtime signal-to-noise. 2) Avoid discomfort by letting people move around, at least periodically, so the body-control circuits stay fluid and discomfort doesn't distract. 3) Every voice is heard by everyone somehow, even without "something to say," because hearing vocalizations is a crucial primate group-calibration protocol.4) Don't decide. Ambiguity and indecisiveness should be preserved as long as possible, because all decisions introduce quantization errors. Long-duration data runs give the best answers.

**What is the optimum way to negotiate?** If two parties intent to negotiate in good faith, the script is simple: Don't box them in. Always leave the other side real decision autonomy, roughly 50%. That doesn't guarantee a settlement, but it does guarantee neither side feels trapped or manipulated, which is the best one could hope for. If both sides maintain that "leave half on the table" symmetry at each stage of a continuous interaction, the result ought to converge as well as possible. The trick to make it work is that compliance has to be self-evident; either side trying to "game" the other is proof of bad faith.

**What is the best way to dance?** This Framework can explain the success of a container for the most general form of dance, called *autonomous motion*. The autonomous motion container requires minimal or no footwear (to stimulate the feet), no talking (to reduce cognitive distractions), and no judgment of self or others (to allow people to move bodies or faces in physically healthful but socially disapproved ways). The resulting "dance journey" then allows a room of people to share a continuous hours-long socio-sensorimotor interaction, each doing what they want in that moment while sharing organic entropy.

**Are economies of attention and of behavior moral hazards in and of themselves?** A "moral hazard" is an economic situation containing an undesired feedback loop, such as a doctor or attorney recommending his own services, or an executive trading on "insider information." An economy of attention is what we have now, in which human attention (often measured in "eyeballs") is attracted by appealing appearance or outright interruption, and its impact sold to promote messages. An economy of behavior is one in which human *behavior* (and not merely exposure to ads) is bought and paid for. The structural problem with both is that those economies literally cannibalize the human nervous system. The better those markets do their jobs, the more distracting and deceptive our environments become, with no obvious upper bound. Perfect market performance would drive everyone crazy.  The moral hazard is that a democracy must discuss the problem of attention markets *inside* those same marketplaces, subject to their incentive structures.

## H7.0 "Paleo everything but violence" provides sensorimotor nutrition

**How does it help me to visit a forest?** For a nervous system evolved to process signals from Nature, Nature provides the best training-data possible. The human nervous system is (among other things) an ultra-sensitive vibration detector and amplifier, using the whole skin-surface along with the ears to detect vibrations. In Paleo times, heat, wind, and vibrations from nearby sources were just as important signals as vision, and those inputs synchronized the skin, ears, and eyes at microtime resolution. Nowadays we fail to get such quality input.  1) Vibrations from machines and gadgets raise the ambient noise levels thousands-fold above natural values. 2) Our skin is covered by clothes, distracting it and muffling incoming vibrations. 3) We spend our time indoors, between walls which create echoes while shielding us from outside sounds. 4) Many sound sources are un-physical, such as Auto-tuned music played through a speaker.  If one stands or sits in a quiet forest, the trees provide both acoustic damping and a source of micro-rustles distributed in three-dimensional space. Animals and birds provide organic sounds, interacting with each other and the wind in organic ways, also distributed in 3-D space nearby. Sights, sounds, smells, and temperature changes are all coherent and coincident at infinite spatial and temporal resolution.

**Should I be thankful for discomfort?**  Yes, most of the time. According to this Framework, most aches and pains are **not** damage to your muscles, but unresolved conflicts in your brain's data-map, leading to quantization-errors about which muscle is tugging exactly where. Your brain doesn't like to make errors, so using those muscle-combinations hurts.  The "thankfulness" part comes in if you produce that discomfort deliberately, as with myofascial self-massage, exertion, or stretching. The temporary, self-inflicted discomfort of pushing your body into a non-damaging sensory situation helps you in two ways. First, you exercise autonomy by choosing the time you feel discomfort, and the amount. Second, the intense sensory stream from focal pressure on a specific zone of ambiguity illuminates that vibrational data-space like no other form of stimulation short of acupuncture. That fresh, clean data gives the brain's self-correcting algorithm what it needs to heal that portion of its control space.

**Does A/B testing damage learning "object-constancy" in children?** One of the hallmarks of childhood cognitive development, the ability to know the world is stable even when you can't see it, may be undermined by a common product-testing process. The process of "A/B testing" is like a clinical trial in medicine, randomly trying one thing or another to see which works best. Online advertisements and software products like Google and Siri routinely randomize parts of their outputs, in order to inform product managers. Any modification at all of an object presented as a single interface, like a home page or digital assistant, is already a violation of the implicit interface contract. A/B testing makes the object no longer a constant object, but a slippery thing not safe for training immature nervous systems. With A/B testing, the changes are entirely random; they are also structured in subtle ways, making the data more pernicious for training. Worst, the "object" being presented and modified will change over time in ways which depend on the user's choices, a correlation which presents the statistical impression that it *cares* about you, exactly the wrong thing for a child to feel about a machine.

**Is oculocentrism a natural orientation?** How much more salient tactile sensation did paleo humans experience than we do now (barefoot, scratchy bushes, naked wrestling)? How much more salient visual sensation do we experience now than then (bright colors, flashing signs etc.)? Is modern life more visually oriented (*oculocentric*) now, compared to feeling-oriented (*propriocentric*) now? Does oculocentrism influence the axis of rotation of the skull? Does it influence balance and posture? Does it lead to a different distribution of spatial awareness than does proprioccentrism? Does it affect the sense of personal autonomy and self-worth? Does oculocentrism lead to pinging?

## H8.0 Humans evolved to resonate ecstatically

**How can I resonate?** Resonating with someone only requires that you be nearby and agree to relax into one another's presence, so both nervous systems can synchronize automatically. This kind of resonance generates high-bandwidth neuromechanical trust, like our trust in our balance. An easy case would be a quiet, intimate gathering, where everyone can see and hear each other at close range. For a more intense experience, two people could stand or lie back-to-back, with spines aligned and touching for several minutes, or stand facing one another in groups of 2-5 with hands one shoulders or the small of the back, breathing or perhaps humming into the space in the middle. The principle is that humans evolved to resonate, so to make it work all you need to do is stop the habits which shut it down.

**How can we make love more contagious than fear?** Love and fear are both legitimate biological reactions, but different: love takes time and resonance, fear is fast and one-way.  Our nervous systems evolved for much more love than fear, so spreading it in person is easy, as with "resonating" above. In fact, absent aversive social conditioning, love should spread among people like wildfire. It actually does, but only in person, and among trusting people, because the back-and-forth microtime signals which convey trust are preserved in proximity.  However, any symbolic interactions, including and especially anything digital, wreck the interaction and microtime trust signals. Unfortunately digital media do convey alerts and warnings extremely well, making them into filters which remove the best human qualities and amplify the worst ones.

**Can mis-tuned instruments tune each other?** Can resonant (e.g. musical) instruments self-tune more effectively in the presence of differently detuned instruments, as with a roomful of mis-tuned pianos? Does the quality or intensity of collective resonance increase with the number of participants? Do those scale with inter-instrument proximity? With the amplitude of vibratory coupling? With its average or peak frequency?  With the narrowness of individual resonances? With reflection from nearby surfaces?

**Can people heal each other with ultrasonic coupling?** What happens when humans exchange ultrasonic vibrations with one another using impedence-matched mechanical coupling, such as a graphite-epoxy bite-bar? Does such high-frequency coupling create the same resonant entrainment as lower?  Are there benefits or dangers not predictable from ordinary sonic coupling? Does ultrasonic resonance feel more intimate than normal physical contact?  Can high-intimacy connections ever be trusted in containers which are incentivized or non-transparent?

**Can human skin feel delicious?** Is that assessment dependent on mood and context?  Is it reciprocal?  Can it be identified and trained?  Can the sensation be "aimed" at a stranger? Is it stronger in groups?  Is it stronger with anonymity? Under ideal circumstances, is sharing anonymous affection healthful?  Under what circumstances might it be risky?  Can such intimate connections be safely shared in untrusted or incentivized environments?

**What would an ideal Yoga or dance studio look like?** Obviously, social comfort critical to enjoying an experience in an embodiment studio.  If that comfort can be maintained, is the experience improved by having mirrors so the peripheral visual system can gather more microtiming data? What about a closer interpersonal distance? Tighter clothing? Less area of clothing?  A wider dynamic range of synchronized exertion and relaxation?

**Can group vibration heal us?** Do humans feel better during and after vibrating in the company of others? Does their health benefit from co-vibration? Does collaborative discussion provide such benefits? Co-singing? Co-vocalizing? Co-breathing? Co-vibrating? Co-touching? Co-gazing? Do the benefits improve with shared intention? With mutual, overt gratitude for the necessary collective effort? Do the people involved feel more connected to one another in proportion to their resonant involvement?

## H9.0 Symmetric spinal health syndrome

**Why does my body become achier and stiffer as I age?** Your body is made of millions of muscle-fibers which vibrate in synchrony. Your brain's primary job is to keep that synchrony perfect, blending microscopic ripples with huge heaves. Perfect vibratory coordination would render your body perfectly sensitive, balanced, and graceful. If that's your life now, you're lucky; most of us have aches, pains, and stiff joints which worsen with age. In this Framework, most of those discomforts trace to data-conflicts in the proprioceptive system, so the ache really is "all in your head." The generic solution is to challenge with stimulation and exertion exactly the locations and configurations which are most uncomfortable, on the theory that discomfort means missing data, and providing data means healing.

**How does my mind work?** A brain is a bandwidth engine, and most of it manages biomechanics, not "mind." Keeping track of right now, the very current moment, takes an enormous amount of processing power, with little left to reconstitute what happened only a few moments ago. In this Framework even "memory" is an evolutionary latecomer, hacked to operate atop a much older continuous-time motor architecture. Furthermore, words and symbols are not only frozen in time like memory, but also involve man-made categorical assignments, so they lie even farther afield from the structures our brains evolved to process. The bad news is, even though we are by far the best species at symbolic cognition, and even though society rewards cognition handsomely, sequential thought moves very slowly, in seconds rather than microseconds. And it moves approximately, since we can't always trust the sources or categories. And it fails to match reality, since the sharp divisions symbolic assignments and categorical divisions depend on don't actually exist in nature. The good news is, our continuous 3-D minds still work blazingly fast on their native turf, if we let them.

**Do data problems cause our itches, aches, & pains?** Do itches occur at locations of temporary, local proprioceptive ambiguity? Does the ambituity resolve when the itch goes away, i.e. when disambiguating data arrives? Are myofascial "trigger spots" locations of persistent ambiguity? Do those resolve and disambiguate with more and deeper data over longer times? Do some sudden musculo-skeletal configuration changes (clicks, pops, snaps) result not from fluid cavitation, but from discontinuous changes in motor strategy? Do some such changes result in sudden pleasure or relaxation? Do other changes result in sudden spasms in newly-activated muscles? Do the textures and sounds of some releases mimic those of physical injuries such as sprains and dislocations?

**Why do people love activities like Yoga so much?** Many spiritual and physical disciplines share common spine-centered and midline-centered practices (including activities not usually thought of as such, like choral singing). Adherents of those disciplines say they benefit approximately according to their participation,  including less overall anxiety, depression, sickness, and discomfort, and better relaxation, sleep, flexibility, balance, physical fluidity, sensory sensitivity, emotional self-regulation, interpersonal relationships, and overall happiness. In this Framework those benefits make sense, because boosting the bandwidth of the central nervous ought to improve the performance of all of its components.

**How can I supercharge my (Yoga/Pilates/dance/pole) practice, or at least feel younger?** First of all, these practices are already very powerful: the more you need them, the more they will challenge you with discomfort. They already provide the benefits of more stable reference frames, smoother and more self-consistent emotional landscapes, and more graceful motion in old age.  They do this by providing hands-on personalized instruction in close proximity, by and in the presence of the same trusted people over extended time.  All such healing is 99% vibration, 1% instruction. Those are features, not bugs, and all are crucial safety nets for the audacious process of rebooting one's own nervous system while continuing to use it for supervising the rebooting process. For such a process to be safe, one absolutely needs affectionate companionship on demand.

If you are in such a lucky situation, here are some new ways to accelerate the process of re-symmetrizing one's nervous system.

1. *Isolate the spine from the periphery.*  One can hang, right-side-up or inverted (by the feet or thighs) to provide lengthwise traction mostly independent of arms and legs. One can  alternate traction with compression, say by doing a headstand. One can remove the influence of the arms and shoulders by supporting the headstand with the *feet*, hands at the side (it's scary!). One can press on the fontanel with smooth or sharp objects, to train it as a locus of muscular constriction and expansion, like the palm. The fontanel operates the central axis of the spine and skull and thus the upper end of your core.
2. *Try acupuncture.*  By injecting body-tremor vibrations into insensitive tissue, an acupuncture needle gives that tissue crucial new data which no exercise could provide. So acupuncture enlivens the nervous system through a complementary, independent channel.
3. *Try "AC grounding" the incisors.*  Weirdly, pressing your incisor-teeth against a solid object  for a few minutes at a time helps straighten the spine too. The key insight comes from Tesla's MECHANICAL THERAPY machine: bodies benefit from ultrasonic vibrations.  Bodies have self-generated ultrasonic vibrations inside themselves already (the ones you sometimes hear as ringing in the ears).  The only thing preventing the body from making sense of the vibrations it already has is the lack of a fixed reference-point. The problem is that even when your bare heels stand on a stone floor, the bones don't touch the stone directly, only through flesh, so they aren't "grounded" at the ultrasonic frequencies they need (a metal bolt from stone to bone might solve the problem, in principle, but that takes surgery). The solution: press the front of the incisor teeth against a hard surface with enough pressure to feel, but not enough to hurt the teeth. Maintaining steady pressure provides the skull a non-vibrating reference signal, what engineers would call "AC grounding," relative to which the spine can now detect and corral its own vibrations.  This process helps relax the jaw and neck, and creates cascade releases elsewhere on the spine.

**Does central spinal activation make me feel better?**  Does intermittently tensioning the spine (e.g. hanging) *while* grounding help anneal the spinal motor map more quickly? Does compressing it (pushing upwards)? Does melodic singing? Vocalizing? Audibly breathing? Feeling sharp pressure centered on the fontanel? Feeling a hard object gently clamped between the molars?

**Do vibrational therapies share common mechanisms?** Does acupuncture recalibrate proprioceptive models by teleporting inertial and body-tremor signals via stiff steel into unresponsive "locked-in" myofascial tissue? Do "sound healings" provide the bodies of participants a sustained 3-D coherent, continuous sound field of simple mechanical origin?  Does such full-spectrum coherence re-synchronize tactile, proprioceptive, and auditory channels?  Do massage-jets and showers provide related vibratory benefits? Does brushing the skin?

**Can I become a hyper-athlete?**  A hyper-athlete has an exceptionally well-tuned and responsive nervous system, not any specific skill or strength.  As yogis and hyper-athletes know, optimal physical activity involves management of attention as much as control of muscles. A focused, rational data-gathering strategy, when applied to re-symmetrizing breath and spine deliberately, has drastic impacts. When the spine and nervous system are both re-symmetrized and managed properly,  they are ideal, and will work as well as any human's could.

**Why do my joints pop?**  Do the discomforts which accompany the transformation include bouts of popping joints, proprioceptive & visual hallucinations, subtle reference-frame shifts, spiritual and intellectual discoveries, sudden fatigue or extra energy, olfactory hyperacuity, digestive problems, heart palpitations, and other spine-associated symptoms which might plausibly have been triggered by local re-aligning of the circuitry for vertebral control?  When a stuck region releases, can it feel and sound like a bone breaking open, but without actual pain?  When a new connection forms, can it feel like a bone collapses?  Do highly salient midline releases near the heart bring co-human feelings?

**Are the muscles I feel inside me impossible hallucinations?** In some sense all sensation is hallucination, since your brain makes it up. You can prove this by looking for sore spots and "trigger points" on your skull. Do they occur disproportionately on prominences or in depressions? Are they hard to minutely localize, i.e. do they correspond to points of proprioceptive ambiguity? Does stimulation engender new sensations in the skull?  Does it improve mobility? Are the sensations felt as muscles, i.e. as stripes of force? Are the sensations limited to the surface of the skull, or are they felt "inside the brain"? Do releases in the skull's cracks (*sutures*) lead to feelings of motion inside them?  Do they lead to sensations that the plates of the skull itself move relative to one another, as if the skull were an elastic solid?  Does feeling the skull bend make it easier to feel the spine twist through the middle of the head?  Does that feel good? Does that feeling pass through the skull's natural geometric axis of rotation?  Does it pass through the location of the pineal gland?

 **Why do people benefit from aerial sports?** Do practitioners of climbing/twisting sports like rock climbing and pole fitness enjoy the same benefits as yogis?  Are their transformations even more spectullar than those of yoga? Does the sport itself employ more spinal tension and elongation? More compression and bending? More torsion? More inversion? Do those specific "postural" exertions bear credit for improving spinal self-awareness, flexibility and fluidity (SSFF)? Does  SSFF bear credit for the benefits?  For the discomforts?

**Should my body feel detailed, or simple?**  You get to choose how to run your body: do you want to go fast, or go fine?  Body control faces the same constraint as Quantum Mechanics:  *There is only so much information in a given block of spacetime.*  To gain resolution in time—the bandwidth goal— you must sacrifice resolution in space.  In this scheme, the fastest data-aggregation geometries (*continuous distributed multiscale operating systems*) must be the simplest shapes. For example, starting from the upper left of the Figure below: Spherical, a one-dimensional breath cycling over time; cylindrical, a one-dimensional spine in space (virtually connected end-to-end); the product space of those two; a bendable spine; and so on, increasing in dimension and complexity. In this scheme, concentrating on one's center re-symmetrizes the whole body.

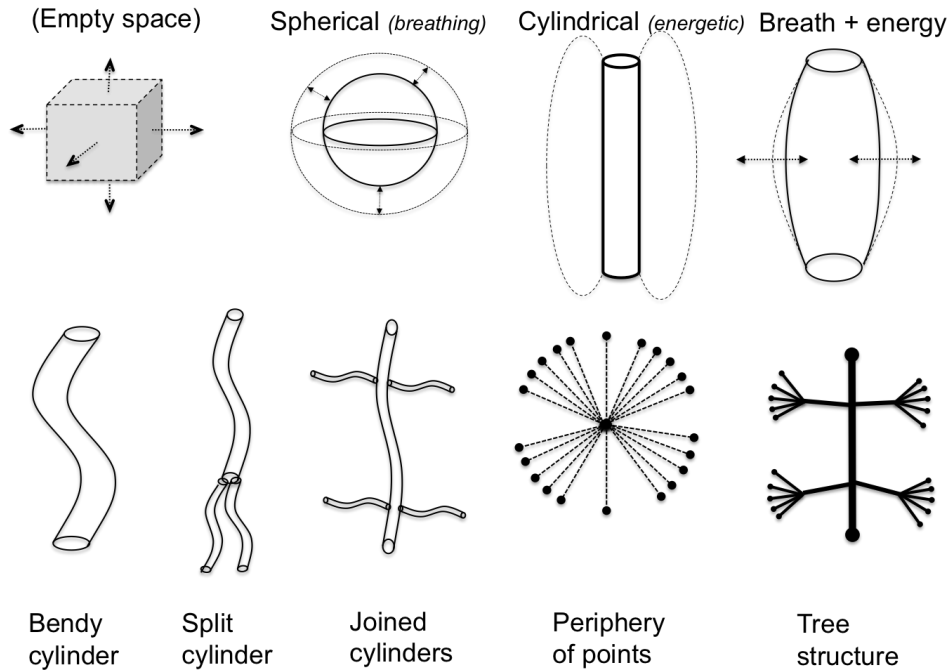

(Empty space)  Spherical *(breathing)*  Cylindrical *(energetic)*  Breath + energy

Bendy cylinder   Split cylinder   Joined cylinders   Periphery of points   Tree structure

**Is the "chakra system" a consequence of spinal software architecture?**
Is the core spine-control architecture built around the neural tube? Are
specific metabolic and musculo-skeletal controls co-located along that
central axis? Would that overlapping architecture create subtle interactions
between local spinal control and metabolic improvements, such as those that
"chakra-like" theories have long claimed?

**Does music ring our spines like violins?** Are the multiscale temporal
patterns of musical overtone series structurally similar to those of chords,
chord progressions, and key changes? Are those in turn similar to the
(slower) structures of polyrhythms? Are those in turn similar to the
structures in vertebrate spines?  Are "Western" instrumentations, harmonies,
and rhythms optimized around those template patterns?  Would signals so
optimized strip from music the human-generated microtime jitter our
nervous systems need?

**Can ultrasonic senses be trained?**  Could hairless skin have evolved to
allow the skin more accurate mechanical signal transduction? Could cave-
people have had an ultrasonic sense, feeling the walls in the dark? Can a
near-naked human body act as a phased-array vibration detector, sensing the
azimuth and elevation of an isolated acoustic source, such as a flying drone,
more accurately than with the ears alone? Do humans in skin-to-skin contact
share ultrasonic signals along with lower-frequency ones? Can people with
supple spines learn to "feel" things near their heads?

**Can you hear the shape of a drum, or of a cave?** Can you "feel" nearby walls with unconscious auditory and ultrasonic senses?  Can your proprioceptive sense "feel" the shape of an extended object you hold?  Can they feel the shape of another person, as a dancer might? Can your incisors feel the shape or size of an object they touch?

## Appendix C. A case for ultra-high-Q spinal resonance

**To first order a human spine is an actively anti-damped resonating cylinder, which means 1) our spines' highest-frequency harmonics are ultrasonic, 2) those harmonics tend resonate with those of others nearby, especially if touching skin-to-skin, and 3) those vibrations converge and tune us collectively, in the same way that pitch and tempo converge among musicians.**

* * * *

A human body is nothing if not an elaborate mechanical system, with the potential (if undamped) to vibrate at over six orders of magnitude of frequency. If there do exist high-frequency, low-amplitude acoustic vibrations inside the skin and myofascial web, those could be used as "carrier waves" underlying tactation, mechanoreception, proprioception, and kinaesthetic sense. Then lower-frequency vibrations, tracked and entrained via those high-frequency carriers, would produce slower but more directly useful gross-motion tasks like balance, walking, jumping, stretching, and breathing. As with modulated radio waves like AM and FM, control of the gross signal depends directly on the quality of the underlying carrier, so a brain's highest-bandwidth task would be to maximize the micro-coherence of muscle activation (*motor spikes*) and travelling acoustic waves in tissue by actively anti-damping motor-spike vibrations.

As in idealization, it is possible one can approximate a vertebrate body like a perfect bell, ringing essentially forever at all frequencies with near-infinite Q. The idea that a hunk of meat can be thought to vibrate perfectly might seem absurd, but the idea is simple.  Suppose that a brain's first-order strategy of managing vibrations amounts to acting like a super-collider control system, with myofascial solitons playing the role of packets of particles. The control system would inject coordinated acoustic waves into tissue (via motor spike timing), and would thereby time subsequent spike firing to "kick" and sharpen that travelling wave as particle-colliders do.  Using such a process, the primary causes of mechanical damping would be eliminated and the flows of energy and entropy would be in equilibrium.  Thus, synthetic very-high-Q behavior could exist, opening the possibility that the very most central and high-bandwidth components of our bodies, our spines, might ring and ring and ring each other like violin-strings sharing bows.

## Appendix D. One validated data point

I would like my own story to serve as an example.

I was a former gymnast and lifelong athlete. At age 50 a chiropractor took an X-ray as I stood up straight. I was shocked to see a calcium-encrusted spine, visibly twisted and dislocated, with hips displaced more than ten degrees. The doctors told me I had arthritis of my entire spine, and measured that my neck rotation and flexion were about half of normal.  Evidently, for decades I had been unaware of a deep spinal disability.

I concluded that my brain's proprioceptive map—the look-up table with which it turns pulses from myofascial mechanoreceptors into a felt sense of which muscles and bones are where and under what stresses—was deeply flawed, and had mis-placed the central axes and locations of key joints, especially in the spine. I had a twisted mental map resulting in *virtual scoliosis*, and an attendant lack of sensation and ability in the crucial "core" muscles.

My zero-parameter theory of my own body is that, originally, some torsional insult to my spine *in utero*—perhaps the umbilical cord twisted around me—allowed my brain to assign a feeling of  "straight" to vertebrae which were in fact mis-aligned, especially in the sternum.  That misalignment of map with muscle did not impact my spine's low-frequency load-bearing capacity (I have always been strong), but it did severely limit my  spine's ability to propagate the ultrasonic myofascial waves which underlie proprioception.  Localized data-conflicts had made my spinal muscles— which ought to be the body's most central, subtle, and sensitive—functionally invisible to my awareness.

This explained many aspect of my life. From childhood until my mid-fifties, I never felt muscular sensation initiating from my spine, so for full spinal flexion or torsion, I had to pull, push, or bend my torso with arms and legs. I always held my breath when concentrating (still do). I can't float in fresh water, even with lungs full. I used mental recollection and triangulation to infer that I had expressed emotions I didn't feel. I use mental imagery more than proprioception to figure how to move. I preferred jump-and-reset activites like rock-hopping to flowing motion.

Over the years since this discovery I have been determined to illuminate my own sensorimotor dark-space with new sensorimotor data, and to learn to operate my musculo-skeletal system as it ought to be.  So far, it has worked just as the Framework says it should. The following encapsulates some aspects of an experience:

> It happens very suddenly, like lightning.  Really. I often hear my "bones" crackle inside my head, a sound very high-pitched, like breaking glass.

Others can even hear it across the room. It can feel like my head was screwed on wrong, then clicks on straight.

Although I know the vibrations emanate from myofascial neck- and skull-control fibers wrapped *around* my skull, the sensation appears, impossibly, *inside* the middle of my brain. In the first place, such hallucination is possible because *all* proprioception (and indeed sensation) is 99.99% hallucinated anyway. But furthermore, the particular texture of my hallucination—feeling muscles where they are not—represents the most efficient use of the brain's native 3-D representational space, and in doing so also lets me feel smooth stripes of force from fontanel to finger and toe, as if my body were not separate bones and muscles, but a single elastic solid, like Gumby.

The glass-break sound has frequencies around 10 kilohertz at least, which means the suddenly-deprecated neuromotor strategy can at once release a slew of fibers within fifty microseconds. That "crack!" made my brain feel open, silent, fluid, and calm at once. What other cause could make a sound so sharp, repeatedly, consistent with internal mechanical sensation, while yielding such immediate relief?